\documentclass{cargese}\usepackage{graphicx}
\usepackage{epsfig}
\let\footnote\savefootnote
\let\footnotetext\savefootnotetext 
\let\footnoterule\savefootnoterule 
\normallatexbib
\newcommand{\eq}{\begin{equation}} 
\newcommand{\eqx}{\end{equation}}
\newcommand{\eqn}{\begin{eqnarray}} 
\newcommand{\eqnx}{\end{eqnarray}}

\newcommand{\f}[2]{\frac{#1}{#2}}
\newcommand{\lra}{\longrightarrow}

\newcommand{\cor}[1]{\left\langle{#1}\right\rangle}

\newcommand{\xpr}{x_\perp}
\newcommand{\sng}{{S_{NG}}}

\newcommand{\var}[1]{\f{\dl \sng}{\dl {#1}}}

\newcommand{\dl}{\delta}

\newcommand{\zx}{z_x}
\newcommand{\wx}{w_x}

\newcommand{\dlgrav}{\dl_{grav}}
\newcommand{\dlb}{\dl_B}
\newcommand{\dlkk}{\dl_{KK}}

\newcommand{\qb}{\overline{q}}
\newcommand{\NN}{{\cal N}}
\renewcommand{\th}{\theta}

\newcommand{\rr}[4]{#1, {\it #2 \/}{\bf #3} #4}

\begin{document}
%------------ article title  ------------------->>

% For a long title use \\ to cut lines.
% In that case, supply  alternate version of the title
% in square brackets, (it will go in the Table of contents during final
% production of the book.
% \articletitle[Short title]{The long version \\ of this title}

\articletitle[Gauge theory scattering]{Gauge theory scattering from\\
the AdS/CFT correspondence\footnote{Talk presented at the NATO ASI
`Progress in String Theory and M-theory', Cargese 1999.}} 

%% optional, to supply a shorter version of the title for the running head:
%%\chaptitlerunninghead{}

\author{Romuald A. Janik}

%% multiple authors at the same institution may be separated with \\
%% like in \author{Samuel Bostaph\\
%%                 and Gregor Kariotis}

\affil{Service de Physique Theorique  CEA-Saclay \\ F-91191
Gif-sur-Yvette Cedex, France\\
and M.Smoluchowski Institute of Physics, Jagellonian University\\ Reymonta
4, 30-059 Cracow, Poland}        
% Your Institution and address. May cut into  separated 
%                 % lines with \\
%
% optional email address
\email{janik@spht.saclay.cea.fr}
%
%%% Repeat the above for multiple authors at different institutions.
%% \author{ }
%% \affil{ }
%% \email{ }

% optional abstract
\begin{abstract}
We use the AdS/CFT correspondence to study near forward scattering of
colourless objects in gauge theory in the high energy limit. We
find an unexpected from the gauge theory perspective `gravity-like'
$s^1$ behaviour of the amplitudes coming from bulk graviton exchange.    
\end{abstract}

%------------ body of article ------------------->>
% Write your article here. 
% Note that the \section{section title}
% command allows for the form \section[short title]{very long\\ title}
% Idem for \subsection and \subsubsection

\section{Introduction}

In this talk we would like to present results on certain aspects of
high energy scattering in gauge theory \index{gauge
theory}\index{gauge theory!high energy scattering} from the point of
view of the AdS/CFT correspondence\index{AdS/CFT correspondence}
\cite{adscft,review}. The full account of these results appeared in 
\cite{adspaper}. The motivation for this study is twofold. Firstly, the
duality between supergravity (string theory) on AdS and
supersymmetric gauge theory allows for a completely new insight into
the dynamics of gauge theory in the strongly coupled
regime. Its consequences regarding the high energy limit of gauge
theory have not been explored so far (see \cite{adspaper,Zahed}).
Secondly, 
expectations on the behaviour of
high energy {\em field theory} amplitudes can 
shed some light on 
the behaviour of the string theory side and 
point out some phenomenae unexpected from the gravity point of
view.
    
The outline of this talk is as follows. First we identify a suitable
observable on the gauge theory side, then we apply the AdS/CFT
correspondence to calculate this quantity and finally we discuss the
implications of our results.

\section{Gauge theory observable}

In this section we will define the gauge theory observable related to
the near foward scattering amplitude:
\eq
\label{e.ampinit}
\f{1}{s} A(s,t)=\int d^2b \;e^{iqb}\;\f{e^{i2\dl(b,s)}-1}{2i}
\quad\quad\quad\quad t=-q^2 
\eqx
where $b$ is the impact parameter (in the following we will denote its
modulus by $L$) and $\dl$ is the phase shift.

In QCD the scattering amplitude
between quarks
in the eikonal approximation (valid in the high energy $s$, fixed (and
small) momentum transfer $t$) is given by a correlation function of
two Wilson lines which follow the classical straight line quark
trajectories \cite{Nacht,VV,Kor}: 
$W_1\lra x_1^\mu=p_1^\mu\tau$ and $W_2\lra
x_2^\mu=\xpr^\mu+p_2^\mu\tau$. The amplitude is in general infrared
divergent and to overcome this it is convenient to consider the
scattering of colourless $q\qb$ pairs of transverse separation $a$. 
In this case the two Wilson lines are replaced by two rectangular
Wilson loops each 
formed by the trajectories of the quark and antiquark forming the
pair. Finally since we will be interested in using the AdS/CFT
correspondence for $\NN=4$ SYM, our quarks and antiquarks will be
substituted in the standard way by (massive) gauge bosons coming from
Higgsing $U(N+1)\rightarrow U(N)\times U(1)$ through the inclusion of
a probe D3 brane. 

\begin{figure}
\centerline{\epsfysize=4cm \epsfxsize=4cm \epsfbox{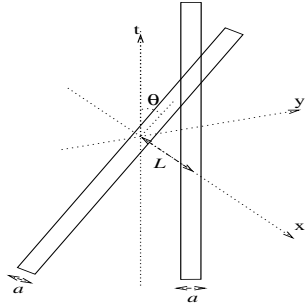}}
\caption{Geometry of the Wilson loops in euclidean space.}
\label{fig}
\end{figure}

As a technical tool we will perform an analytical continuation to
euclidean space. Thus we consider Wilson loops\index{Wilson loop}
which form an angle $\theta$ in the longitudinal $t$-$y$ plane. The
`width' of the loops is along the transverse direction '$x$' (see
figure \ref{fig}).  

Explicitly we have to compute
\eq
-2i \int d^2\xpr e^{iq\xpr} \cor{\f{W_1W_2}{\cor{W_1}\cor{W_2}}-1}
\label{e.observable}
\eqx
as a function of the angle $\th$, and
after performing the calculation we analytically
continue
$\th \lra -i\chi\sim -i \log\left(\f{s}{2m^2}\right)$
to obtain (\ref{e.ampinit}).  
In the next section we apply the AdS/CFT correspondence to calculate
(\ref{e.observable}).

\section{Supergravity calculation}

According to the now standard prescription, a Wilson loop on the gauge
theory side corresponds to a minimal surface string worldsheet in AdS
bounded by this loop \cite{Wilson}. The correlation function of two
Wilson loops\index{Wilson loop!correlation function of two Wilson loops}, in 
the case when their separation is much larger than their transverse
sizes, will be given by the exchange of supergravity fields
between the two associated minimal string worldsheets. 
The calculation is technically completely analogous to the one of
\cite{MaldCor}, the only change is a different $\th$-dependent
geometry of the loops, and the resulting interpretation as a gauge
theory scattering amplitude. Due to the analytical continuation
it turns out that the hiearchy of importance of
various supergravity fields is different from that in \cite{MaldCor}.

The calculation involves two steps:\\
(1) determination of the couplings of the various supergravity
fields\index{supergravity fields}
to the string worldsheet,\\
(2) calculation of the interaction using the appropriate bulk to bulk
Green's function.\\
The couplings are found by varying the Nambu-Goto action with respect to
the background supergravity fields. The resulting contribution of the
relevant field to the gauge theory scattering phase shift is then
given by
\eq
\label{e.master}
\!\!\!\! \!\! \f{\cor{W_1W_2}}{\cor{W_1}\cor{W_2}}=\exp\left
( \f{1}{4\pi^2{\alpha'}^2} \int 4 dt_1 
dt_2 \f{dz dw}{\zx \wx} \var{\psi}(t_1,z) G(x,x')
\var{\psi}(t_2,w) \right)
\eqx
where $\psi$ denotes a generic supergravity field.  
Using the symmetry properties of the AdS geometry one can show that
the whole $\th$- (and consequently energy $s$-) dependence is encoded in
the couplings. All the detailed properties of the Green's function
only change the $a/L$ dependence of the amplitude.

We performed the calculations for the tachyonic scalars $s^I$,
lightest modes of the dilaton, of the
2-form field and of the graviton. Similarly as in flat space  
the energy dependence turns out to be related to the spin of the exchanged
particle. 

The phase shift coming from the scalars behaves like $\dlkk\sim
(a/L)^2 \cdot (1/\sin\th) \sim (a/L)^2 \cdot s^{-1}$. From the 2-form field 
$\dlb\sim (a/L)^4 \cdot (\cos\th/\sin\th) \sim (a/L)^4 \cdot s^{0}$ and for the
graviton we get $\dlgrav\sim (a/L)^6 \cdot (\cos^2\th/\sin\th) \sim
(a/L)^6 \cdot s^1$. Thus the graviton dominates at high energies.

We would like to comment on the
limitations of the above calculation. We considered only
single particle exchange. This is
legitimate in the large $N$ limit but for high enough energies
may be inadequate. In fact analytically continuing to Minkowski signature
within the AdS space one can check that the gravitational field
produced by one of the worldsheets can be considered as a small perturbation
around the background geometry only for $s^2 \ll (L/a)^7$.  

The surprising result from the gauge theory side is the $s^1$ dependence
of the graviton contribution. In field theory we do expect such higher
powers of $s$ to be supressed by unitarity corrections, but it is
a difficult challenge to envisage how this would look like
on the supergravity side, where there is no obvious process cancelling
the attractive graviton-mediated interaction at {\em high energies}.  
Probably something very specific to string theory on AdS would have to
intervene.

%------------ end of article ------------------->>

%% optional
\begin{acknowledgments}
I would like to thank the organizers and participants of the Cargese
ASI for a very stimulating school, and Robi Peschanski with whom
these results were obtained.
This work was supported in part by KBN grants 2P03B08614 and 2P03B01917.
\end{acknowledgments}

%% appendix optional
%\chapappendix{This is the Appendix Title}
%This is an appendix with a title.

%\chapappendix{}
%This is an appendix without a title.

%

\end{document}